\documentclass[aps,prd,amssymb,amsmath,nofootinbib]{revtex4}
\usepackage[dvips]{graphicx}
\begin{document}
\title{Clock rate comparison in a uniform gravitational field 
\footnote{Present paper contents were deeply analyzed in a more
recent paper of the same author: arxiv.org gr-qc/0503092.
The new paper has some differeces in the problem approach,  
and contains different extra material, expecially about metric 
porperties. It also contains all the results and proofs of the
present paper. For this reason we strongly suggest to new readers
to avoid the lecture of this paper and to jump to the new one.
}}
\author{Marco Alberici}
\affiliation{Dipartimento di Fisica, Universit\`a di Bologna 
Via Irnerio~46, 40126 Bologna, Italy.}
\email[]{alberici@bo.infn.it}
\altaffiliation{I.N.F.N. member, Sezione di Bologna.}
\date{\today}
\begin{abstract}
A partially alternative derivation of the expression for 
the time dilation effect in a uniform static gravitational field
is obtained by means of a thought experiment in which rates 
of clocks at rest at different heights are compared using as
reference a clock bound to a free falling reference system 
(FFRS). Derivations along these lines have already been proposed,
but generally introducing some shortcut in order to make the 
presentation elementary. The treatment is here exact: the clocks
whose rates one wishes to compare are let to describe their 
world lines (Rindler's hyperbolae) with respect to the FFRS, and
the result is obtained by comparing their lengths in space-time. 
The exercise may nonetheless prove pedagogically instructive
insofar as it shows that the exact result of General Relativity (GR)
can be obtained in terms of physical and geometrical reasoning
without having recourse to the general formalism.
Only at the end of the paper the corresponding GR metric is derived, 
to the purpose of making a comparison to the solutions of Einstein
field equation.
This paper also compels to deal with a few subtle points inherent in 
the very foundations of GR.
\end{abstract}
\maketitle
\newcommand{\be}{\begin{eqnarray}}
\newcommand{\ee}{\end{eqnarray}}
\section{Introduction}
It is possible to show that the GR redshift formula for a uniform 
gravitational field
\be 
\label{eq.price} \frac{\Delta \nu}{\nu}=\frac{gh}{c^2}, 
\ee
it is an exact consequence of the EP, and its derivation does not require
GR formalism \cite{price}. As is well known, formula (\ref{eq.price}) was 
experimentally verified for the first time in 1960 by Pound and Rebka
\cite{pound}.
\par GR is a well-established theory, but it very often happens that
its application to some specific case involves subtle points on which
the agreement is not general: as a result, the conclusions of the analysis
reported are often not univocal. This is the case, in particular,
for the object of the present article, namely redshift in a uniform 
gravitational field.
\par In Weinberg's treatise \cite{wein}, for instance, the gravitational 
redshift effect is thus commented: `For a uniform gravitational field, this 
result could be derived directly from the Principle of Equivalence, 
without introducing a metric or affine connection'.
\par We have in fact already recalled that such derivations are
possible and in an exact way \cite{price}. Independently of this
conclusion, one may always introduce transformations between reference
frames at rest with respect to the matter generating the field and
FFRS which, in the case of a uniform field, can become extended, or
global. The most natural way to describe such a transformation are
Rindler hyperbolae. But Rindler himself \cite{rindler}, at the end
of the paragraph of his book in which they are introduced and analysed,
and after discussing the fact that SR can deal in a proper way with 
accelerated frames, states: `A uniform gravitational field can
not be constructed by this method'. 
About 15 years ago, E.A. Desloge \cite{des1,des2,des3,des4}, in a sequence
of four articles, carried out a systematic investigation on uniformly
accelerated frames and on their connection with gravitation.
In agreement with Rindler's statement Desloge concludes that there is no exact 
equivalence between FFRF and uniform gravitational fields \cite{des3,des4}.
By contrast, only three years later on this journal, E. Fabri \cite{fabri} 
derived the expression of the time delay in a uniform gravitational 
field by explicit use of a FFRF, from which the world lines of objects at
rest with the masses that generate the field are seen as Rindler hyperbolae.
As far I know, Fabri's paper is still the only one approaching 
the problem in this way.
\par In order to see whether a reconciliation between these apparently 
conflicting views could be obtained, I tried to set up an ideal
experiment in which the rates of two standard clocks 
\footnote{We will not come back here to the meaning to be attributed
to expressions such as `Standard Clock' or `Standard Rods', as the question 
has already been discussed in many books and papers and also in one of Desloge's
articles. We will just compare the proper time elapsed in the two RF.}
sitting at different heights in a uniform and static gravitational field
are compared. I analysed in detail the measuring process as seen from a FFRF,
chosen in the only way that does not lead to physical contradictions. 
The choice of the FFRF and the final results are in agreement with Fabri's.
\section{Free-falling reference systems}
\label{acc}
EP is the basic idea from which the bulk of GR theory was developed. 
In our applications we will not deal with charged bodies and with
non-gravitational forces in general. We are therefore entitled to 
overlook differences between weak and strong EP, differences that are
present only when non-gravitational interactions are considered. 
We just consider the EP in this form:
\begin{quote}
{\em A body in free fall does not feel any acceleration, 
and behaves locally as an inertial reference system of SR.}
\end{quote}
We know that physical gravitational fields are not uniform, 
and there always exist tidal forces that oblige to consider the 
validity of the EP only locally in space and time. 
The ideal case of a uniform and static field is however worth considering 
for its conceptual interest. In this ideal case FFRS, which in strictly 
physical situations are local inertial systems, would become non-local.
\par EP invites to consider FFR systems the proper inertial systems. 
Let us then try to build a non-local FFRF, denoted as $K$, in a 
2-dimensional space-time universe and study the world line of a particle 
at rest with the masses that generate a uniform gravitational
field $g$.  Since this world line is considered with respect to the FFRF, 
we will associate to the particle a (local) frame $\Sigma^-$.
It is easily proved \cite{rindler} that the world line of a body falling 
along the vertical direction $x$ with constant proper
acceleration $g$ is described by the hyperbola of equation:
\be \label{eq.rin} x^{2}-c^2t^{2}=\frac{c^4}{g^{2}}. \ee

Since we are considering  acceleration with respect to $K$, it is 
material points at rest in  $\Sigma^-$ that will describe an hyperbola 
of Eq.~(\ref{eq.rin}) in the Minkowski space-time of $K$ (with an 
acceleration pointing in the direction opposite to gravity). 
At $t=0$ an observer sitting in the FFRF and one at rest with the source of the
gravitational field (sitting in $\Sigma$) have the same instantaneus rest frame and 
measure the same acceleration value (the first observer by observation of $\Sigma$'s
trajectories, the second one measuring the force acting on him). 
On the other hand, an observer sitting $\Sigma^-$ will see material points 
at rest in $K$ describe more complicated world lines. 
They have been explicitly calculated \cite {hamilton}, and turn out indeed 
to be very different from those described by Eq.~(\ref{eq.rin}) 
(on the same subject see \cite{ror}).
Fortunately we do not need to deal with them in order to solve our problem.\\
In the following $x$ will be called `coordinate position' and $t$ 
`coordinate time', in order to distinguish them from proper coordinate and
proper time measured by $\Sigma^-$. As usual $c$ is the speed of light, 
but for sake of simplicity in the following we will adopt units with
$c=1$.
\par The space coordinate at time $0$ is $x_0=1/g$, so that Eq.~(\ref{eq.rin}) 
becomes
\be 
\label{eq.rin2} x^{2}-t^{2}=x_0^2. 
\ee
A second RF $\Sigma ^+$ which, at $t=0$, lies higher than $x_o$  by the 
amount $h$ as measured by $K$ is described by the following hyperbola
\be \label{eq.rioo} x^{2}-t^{2}={(x_{0}+h)}^{2}, \ee
where now gravitational acceleration depends on the coordinate. 
Because of this dependence we introduce the following notations:
\be g^+=\frac{1}{x_{0}+h} \qquad \qquad g^-=\frac{1}{x_{0}}.\ee
The fact that the acceleration depends on the coordinate value, seems to be 
in contradiction with the uniformity of the field. We will discuss this 
problem in section \ref{space} where we shall see that our choice is the 
only possible one. In this section, we shall prove that adopting 
Eq.~(\ref{eq.rin2}) and Eq.~(\ref{eq.rioo}), 
three important physical conditions are satisfied:
\begin{itemize}
\item[(1)] Local acceleration does not depend on time;
\item[(2)] The ratio between proper time intervals at different altitudes 
is also constant in time (possibly depending on the frames altitude), otherwise 
a Pound and Rebka-like experiment would give every day a different result;
\item[(3)] Radar distance between two observers situated at a different height 
is a constant quantity which is time independent (it will as well possibly 
depend on the altitude).
\end{itemize}
\par Properties (1), (2) and (3) are all experimentally verified. 
Property (1) is surely satisfied because it is the basis for deriving
Rindler motion equation \cite{rindler}. In section \ref{proper} 
and \ref{light} we will prove also property (2) and (3)
All this properties were already proven by desloge \cite{des2},
for the case of uniformly accelerated frames. We will prove this
properties by a direct calculation that seems to be more intuitive and 
easy to understand.
\subsection{Proper time intervals}
\label{proper}
We need to calculate the proper time interval along a world line 
between two generic coordinate times $t_1$ and $t_2$:
\be \tau_{12} = \int_{t_1}^{t_2}\sqrt{1-V^{2}}dt, \ee
where $V$ it is the coordinate velocity in units with $c=1$. Differentiating 
Eq.~(\ref{eq.rin2}) or Eq.~(\ref{eq.rioo}) one obtains $V =t/x$
and using Eq.~(\ref{eq.rin2}) and Eq.~(\ref{eq.rioo}) in order to eliminate 
$x$, we can calculate the proper time interval $\tau^-$ along the
lower and $\tau^+$ along the upper hyperbola
\be
\label{eq.tau0}
\tau^-_{12} = \int_{t_1^-}^{t_2^-}\frac{1}{\sqrt{1+(\frac{t}{x_0})^2}}dt
=x_0\int_{\frac{t_1^-}{x_{0}}}^{\frac{t_2^-}{x_{0}}}\frac{1}{\sqrt{1+y^{2}}}dy
\ee
\be
\label{eq.tau+}
\tau^+_{12} =\int_{t_1^+}^{t_2^+}\frac{1}
{\sqrt{1+(\frac{t}{x_0+h})^2}}dt
=(x_0+h)\int_{\frac{t_1^+}{x_{0}+h}}^{\frac{t_2^+}{x_{0}+h}}
\frac{1}{\sqrt{1+y^{2}}}dy. 
\ee
\begin{figure}[hbp]
\begin{center}
\includegraphics[scale=0.22]{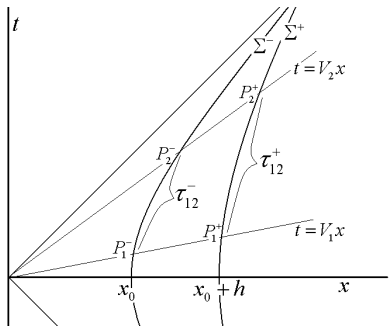}
\caption[short TOC caption]{\em Proper times in $\Sigma^+$ and $\Sigma^-$}
\label{fig.clockdelay}
\end{center}
\end{figure}
Considering the intersection of the two hyperbolae with two lines 
of constant coordinate velocity (see Fig.~\ref {fig.clockdelay})
\be
t=V _1 x \qquad t=V _2 x \qquad \qquad
|V_1|,|V_2|<1,
\ee
one obtains two couples of points $P=P(x,t)$ in the Minkowski 
diagram, whose coordinates are:
\be 
P_1^-= \frac{\gamma _1}{g^-}\big (1,V_1 \big)  \quad
P_1^+= \frac{\gamma_1}{g^+} \big (1,V_1 \big ) 
\ee
\be 
P_2^-= \frac{\gamma _2}{g^-}\big (1,V_2 \big)  
\quad P_2^+= \frac{\gamma _2}{g^+} \big (1,V_2 \big ), 
\ee
where we have introduced the usual notation $\gamma=1/ \sqrt {1-V^2}$.
Coordinates of $P^-$ and $P^+$ are in simple proportions, 
and we have in particular
\be
\frac {t_1^+}{t_1^-}=\frac {t_2^+}{t_2^-}= \frac{g^-}{g^+}=\frac {x_0+h}{x_0},
\ee
that proves the equality of integration extremes defining 
$\tau_{12}^+$ and $\tau_{12}^-$. Dividing Eq.~(\ref{eq.tau+}) by
Eq.~(\ref{eq.tau0}), one finds:
\be
\label{eq.rap} \frac {\tau_{12}^+}{\tau_{12}^-}
=\frac{g^-}{ g^+}=\frac{x_0+h}{ x_0},
\ee
so that one can conclude that:
\begin{quote}
{\em Proper times intervals along two Rindler hyperbolae which are seen 
under the same angle centered in the origin are in a fixed proportion.}
\end{quote}
Considering as simultaneous all events laying on straight lines through 
the origin, the above statement proves property (2). 
This simultaneity will be proven in the next section.
\par Considering that $x_0=1/g^-$, Eq.~(\ref {eq.rap}) can also be 
recast in the following form:
\be 
\frac {\Delta \tau}{\tau}=\frac
{\tau_{12}^+-\tau_{12}^-}{\tau^-_{12}}=g^-h.
\ee
\subsection{Radar distance and light signals between Rindler hyperbolae}
\label{light}
\begin{figure}[hbp]
\begin{center}
\includegraphics[scale=0.21 ]{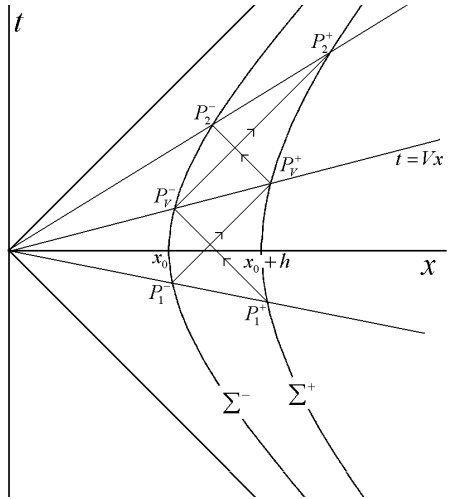}
\caption[short TOC caption]{\em Light signal between two hyperbolae} 
\label{fig.aligned}
\end{center}
\end{figure}

Let us draw a generic line through the origin in Minkowski space, 
which identifies points having the same coordinate velocity:
\be
t=V x \qquad \qquad  |V|<1.
\ee
Intersecting this line with the low and upper hyperbolas identifies 
the points
\be 
P_V^-=\frac{ \gamma}{g^-} \big (1,V \big);\qquad \quad
P_V^+= \frac{\gamma}{g^+} \big (1,V \big). 
\ee
The equations of light rays departing from $P_V^-$ and $P_V^+$ 
are respectively
\be 
t=x-\gamma \frac {1-V}{g^-}; \qquad t=-x+\gamma \frac{1+V}{g^+}. 
\ee
Solving the system together with the Hyperbola equations 
(\ref{eq.rin2}) and (\ref{eq.rioo})
\be
\begin{cases}
&t=x-\gamma \frac {1-V}{g^-}\\[5pt]
&x^{2}-t^{2}= \frac{1}{(g^+)^2}
\end{cases}
\hspace{2cm}
\begin{cases}
&t=-x+\gamma \frac{1+V}{g^+}  \\[5pt]
&x^{2}-t^{2}=\frac{1}{(g^-)^{2}}
\end{cases},
\ee
leads to the following intersection points:
\be
 P_2^- \!\! = \!\! \frac{\gamma g^+}{2} \!
\Big ( \! \frac{1+V}{(g^+)^2} \! + \! \frac{1-V}{(g^-)^2},
\frac{1+V}{(g^+)^2} \! - \! \frac{1-V}{(g^-)^2} \! \Big ); 
\quad
 P^+_2 \!\! = \!\! \frac{\gamma g^-}{2} \!
\Big ( \! \frac{1+V}{(g^+)^2} \! + \! \frac{1-V}{(g^-)^2}, 
\frac{1+V}{(g^+)^2} \! - \! \frac{1-V}{(g^-)^2} \! \Big ).
\ee
%
%
%
In the same way it is possible to calculate intersection points 
between the two hyperbolas and the incoming light rays:
\be
 P_1^- \!\! = \!\! \frac{\gamma g^+}{2} \!
\Big ( \! \frac{1+V}{(g^-)^2} \! + \! \frac{1-V}{(g^+)^2} ,  
\frac{1+V}{(g^-)^2} \! - \! \frac{1-V}{(g^+)^2} \! \Big ); 
\quad
P^+_1 \!\! = \!\! \frac{\gamma g^-}{2} \!
\Big ( \! \frac{1+V}{(g^-)^2} \! + \! \frac{1-V}{(g^+)^2}' 
\frac{1+V}{(g^-)^2} \! - \! \frac{1-V}{(g^+)^2} \! \Big ).
\ee
%
%
%
For any value of $V$, we have the following proportions
\be
\frac{x_1^+}{x_1^-}=\frac{x_2^+}{x_2^-}=\frac{t_1^+}{t_1^-}=
\frac{t_2^+}{t_2^-}=\frac{g^-}{g^+}=\frac{x_0+h}{x_0}, 
\ee
from which one see that $P_1^-$ is aligned with $P^+_1$, and $P_2^-$ 
is aligned with $P^+_2$ (as showed in Fig.~\ref {fig.aligned}).

We have thus found the following result which, as far as we 
know, was not previously pointed out:
\begin{quote}
{\em Light rays through a couple of events with the same coordinate 
velocity laying on a Rindler hyperbola, intersect other Rindler
hyperbolae in a couple of points that have in turn the same 
coordinate velocity}.
\end{quote}
This property it is an extra bonus, as we did not request it in 
our premise at the beginning of section \ref {acc}, and it will make it 
easier to perform clock rate comparison in an operational way.
As a result of the above calculations, it is clear that observers in 
$\Sigma^-$ and in $\Sigma^+$ will consider simultaneous all the events
laying on straight lines through the origin of the Minkowski diagram. 
With this definition of simultaneity we can calculate the proper time
$\tau_S^-$ elapsed in $\Sigma^-$ when a light ray covers the distance 
necessary to go from $\Sigma^-$ to $\Sigma ^+$ and the proper time
taken for the return trip:
\be  
\tau_S^- =\int_{t^-_V}^{t^-_2}\frac{1}{\sqrt{1+(\frac{t}{x_0})^2}}dt=
x_0\int_{\frac{t^-_V}{x_{0}}}^{\frac{t^-_2}{x_{0}}}\frac{1}{\sqrt{1+y^{2}}}dy
\ee
\be 
\tau_R^- =\int_{t^-_1}^{t^-_V}\frac{1}{\sqrt{1+(\frac{t}{x_0})^2}}dt=
x_0\int_{\frac{t^-_1}{x_{0}}}^{\frac{t^-_V}{x_{0}}}\frac{1}{\sqrt{1+y^{2}}}dy.
\ee
The calculation leads to constant quantities \footnote
{If we tried instead to calculate the coordinate $t$ time elapsed 
during the trip between $\Sigma^-$ and $\Sigma^+$, we would find the
following expression $ t=\gamma \frac {h(2x_0+h)}{x_0+h}$, which 
is not constant due to its dependence on $\gamma$.}

\be 
\label{eq.radar-} 
\tau_S^-=\tau_R^-=x_0\ln{\frac{x_0+h}{x_0}}=\frac {1}{g^-}\ln {\frac{g^-}{g^+}}.
\ee
Calculation of proper times relative to an observer at rest with 
$\Sigma^+$ leads also to an equal and constant time,
whose value is in this case:
\be 
\label{eq.radar+}
\tau ^+_S=\tau^+_R= (x_0+h)\ln
{\frac{x_0+h}{x_0}}. 
\ee
This is in our view an important result, which, together with our new 
notion of simultaneity between $\Sigma^-$ and $\Sigma^+$ proves the
following statement:
\begin{quote}
{\em The proper time, as measured by an observer on a Rindler hyperbola, 
taken by a light pulse to reach another Rindler Hyperbola is equal to the 
proper time taken for the return trip, and this proper time is constant.}
\end{quote}
With the proof of this property, we completely fulfilled request (3) 
of section \ref{acc}.
\section{Measurements}
\label{measures}
\subsection{Clock delay}
\label{mis}
All necessary physical requirements having thus been satisfied, we 
feel entitled to set up a thought experiment in which two identically
constructed clocks sit at rest, in a uniform and static gravitational 
field, at fixed distances along a field line of force. As has become
customary, we will give a name to the physicists involved in the 
measuring processes; let's say that Alice enjoys her life staying at rest
with $\Sigma^-$ and Bob staying higher in $\Sigma ^+$. They will perform 
a measure in the following way (see Fig.~\ref{fig.aliceBob}):
\begin{figure}[hbp]
\begin{center}
\includegraphics[scale=0.24 ]{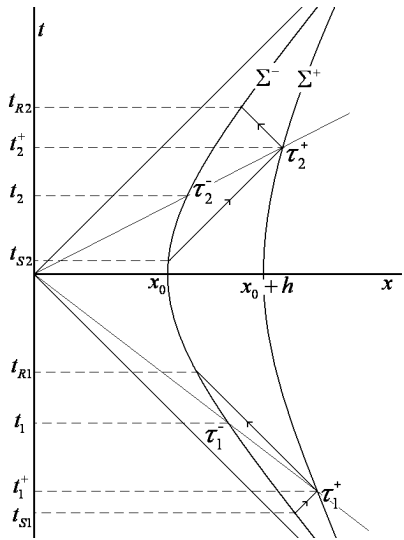}
\caption[short TOC caption]{\em Alice and Bob compare their clock rate} 
\label{fig.aliceBob}
\end{center}
\end{figure}

\begin{description}
\item[a)] at time $t_{S1}$ Alice sends a light signal upwards;
\item[b)] at time $t^+_1$ Alice's light pulse is received by Bob. 
When Bob receives the signal he starts his clock and immediately
sends back downwards a return light signal;
\item[c)] at time $t_{R1}$ Alice receives the return signal from Bob. 
Adopting Einstein clock synchronization, she will argue that Bob clock
started at the intermediate time between sending and receiving the light pulse. 
According to her clock this time is $\tau_1^-=\frac{\tau_{S1}^-+\tau_{R1}^-}{2}$, 
and corresponds to coordinate time $t_1$. In the previous section we saw that, 
with respect to the origin, this event is aligned with the event in which 
Bob started his clock;
\item[d)]
at time $t_{S2}$ Alice sends a light signal upwards;
\item[e)] at time $t^+_2$ Alice light pulse is received by Bob. 
When Bob receives the signal he stops his clock and immediately
sends back to Alice a return signal, in which the information of 
the proper time elapsed in Bob's clock is stored;
\item[f)] at time $t_{R2}$ Alice receives the return signal from Bob. 
She will argue that Bob clock had stopped at the intermediate time between 
sending and receiving the light pulse. According to her clock this time is
$\tau_2^-=\frac{\tau_{S2}^-+\tau_{R2}^-}{2}$ which correspond to coordinate 
time $t_2$.  We have again that in the Minkowski space-time of $K$, this 
event is aligned with the origin and the event in which Bob stopped his clock.
\end{description}
The necessary measurements have thus been done, and Alice can compare her 
time with Bob's. It is immediately seen that any dependence on
starting and stopping times cancels out, and that she will be able to compare 
two proper times seen under the same angle centered in the
origin: $\tau^-_{12}=\tau^-_2-\tau^-_1$;  $\tau^+_{12}=\tau^+_2-\tau^+_1$.
These proper times satisfy Eq.~(\ref{eq.rap}) and Alice concludes~
\footnote{The experiment could be also done by sending starting and stopping 
signals from $\Sigma^+$. In this case Bob would draw the very same
conclusion as Alice and they would both agree on the fact that Bob's 
clock is faster than Alice's. It worth stressing that this situation is
different from that arising in SR with respect the time dilation 
effect, which is observed in a symmetric way by the two observers involved.
} that
\be
\label{eq.dilat}
\frac{\Delta \tau}{\tau}=\frac{\tau^+_{12}-\tau^-_{12}}{\tau^-_{12}}=g^-{h}.
\ee
It is also possible to compare Bob's proper time $\tau^+_{12}$ with 
Alice's calculated either between the two emission or reception events.
This would not make any difference, because in both cases we would 
have only added to and then subtracted from $\tau^-_{12}$ the same
quantity (the times taken by the outward and return trips given by 
$\tau _S^- = \tau _R^- = x_0 \ln{\frac {x_0+h}{x_0}}$).
\subsection{Space dilation}
\label{spacedil}
In Eq.~(\ref {eq.radar-}) and Eq.~(\ref{eq.radar+}) we calculated 
the proper time elapsed in $\Sigma^-$ and $\Sigma^+$ for sending and
receiving a light ray. Multiplying this time by the speed of light 
($c=1$ in our units) we have radar distances $h^-$ (measured by
$\Sigma^-$) and $h^+$ (measured by $\Sigma^+$)
\be
\label {eq.dist}
 h^-=x_0\ln {\frac {x_0 +h}{x_0}}; \qquad
 h^+=(x_0+h)\ln{\frac {x_0 +h}{x_0}},
\ee
%
%
whose ratio give the following expression analog to time dilation:
\be 
\label{eq.distanceratio}
\frac{h^+}{h^-}=\frac{x_0+h}{x_0}=\frac{g^-}{g_+}.
\ee
Keeping in mind that $x_0=1/g^-$ and $x_0+h=1/g^+$, solving 
Eq.~(\ref{eq.dist})  for $h$ one gets
\be h=x_0(e^{g^-h^-}-1)=x_0(e^{g^+h^+}-1), \ee
which makes it possible to rewrite the time dilation formula 
using local coordinates:
\be 
\label{eq.rinexp} \frac{
\tau^+}{\tau^-}=1+g^-h=e^{g^-h^-}=e^{g^+h^+}. 
\ee
One can prove that Eq.~(\ref{eq.rinexp}) is consistent with 
the following general result of GR \cite{rindler2},
\be 
\label{eq.rinorig}
\frac {\nu^+}{\nu^-}=e^{ \phi(x_0+h)-\phi(x_0)},
\ee
which expresses the redshift in terms of the Newtonian potential 
$\phi$ of the gravitational field. Indeed, calculating the potential
difference we find:
\be
\phi (x_0+h)-\phi (x_0)=\int_{x_0}^{x_0+h}g(x)dx
=\ln \frac{x_0+h}{x_0}=\ln (1+g^-h)
\ee
which, inserted in Eq.~(\ref{eq.rinorig}), reproduces our 
time dilation formula (\ref{eq.rinexp}).
\par  A different expression was obtained by Desloge \cite{des4} 
who considered Eq.~(\ref{eq.rinexp}) valid only for a
uniform accelerated frame. For uniform gravity fields, 
Desloge proves the following expression
\be 
\label{eq.desloge}
\frac{\nu^+}{\nu^-}=e^{gh}, 
\ee
where gravitational acceleration is the same for top and bottom 
observers and the distance is measured from a FFRF. Eq.~(\ref{eq.desloge}) 
is also consistent with Eq.~(\ref{eq.rinorig}), but its derivation is 
based on a particular metric:
\be
\label{eq.metricaexpo}
ds^2=e^{2gh}dt^2-dh^2,
\ee
whose exact determination involves solving Einstein's field equation
\cite{rindler3}, and does not lead to a uniquely defined result \cite{ror}. 
We will show in section \ref{metric} that the metric of Eq.~(\ref{eq.metricaexpo}) is
correct but the distance coordinate must be interpreted as radar distance 
measured from an observer at rest with the field.
\par For a better understanding of space dilation it is useful to calculate 
the infinitesimal distance measured with radar methods by two observers 
at rest with the field at different heights. We will use $dh^-_x$ for 
denoting radar measurements of an infinitesimal length located at coordinate 
$x$, performed by an observer in $\Sigma^-$, and $dh^+_x$ if the observer is 
located at $\Sigma^+$. A straightforward calculation, at first order, gives:
\be
\label {eq.delta-} 
dh^-_x=x_0 \bigg( \ln{\frac{x+dh}{x_0}}-\ln {\frac{x}{x_0}}\bigg)
\approx \frac{x_0}{x}dh
\ee
\be 
\label {eq.delta+}
dh^+_x=(x_0  + h) \bigg ( \ln{\frac{x+dh}{x_0}}  -  
\ln {\frac{x}{x_0}}  \bigg) \approx \frac{x_0+h}{x}dh, 
\ee
whose ratio is:
\be 
\label{eq.deltah} 
\frac {dh^+_x}{dh^-_x}=\frac{x_0+h}{x_0}=\frac{g^-}{g^+}. 
\ee
\par An important consequence of Eq.~(\ref{eq.deltah}) is that 
$\Sigma^-$ and $\Sigma^+$ measure the same infinitesimal work per unit mass:
\be
d \phi =g^+dh^+=g^-dh^-.
\ee
\subsection{Gravitational acceleration}
\label{gravityacc}
In this subsection, we want to compare physical measurements 
performed by different observers in the same local frame.
Let us consider two different measurements in Bob's frame ($\Sigma^+$):
\begin{description}
\item [1] Bob ($\Sigma^+$) performs direct measurements with clocks 
and rods built in $\Sigma^-$ (or built in $\Sigma^+$ but with the very
same physical procedure as those used in $\Sigma^-$)
\item [2] Alice ($\Sigma^-$) observes Bob's frame by sending and 
receiving reflected light pulses.
\end{description}
The two measurements are in the following relations
\be
\label{eq.proportion}
\underbrace{dh_{x_0+h}^-}_{Alice's~measure}=\frac{g^+}{g^-} 
\underbrace{dh_{x_0+h}^+}_{Bob's~measure};\qquad \quad
\underbrace{d\tau_{x_0+h}^-}_{Alice's~measure}=\frac{g^+}{g^-}
\underbrace{d\tau_{x_0+h}^+}_{Bob's~measure}.
\ee
Therefore, for the same couple of events, Alice obtains the same 
average velocity measured by Bob, but measuring shorter space and
time intervals. Because of these contractions, it easy to prove that 
Alice will measure larger accelerations than those measured by Bob:
\be
\label{eq.gratio} \underbrace{g_{x_0+h}}_{Alice's~measure}=
\frac{g^-}{g^+}\underbrace{g_{x_0+h}}_{Bob's~measure}.
\ee
We know that the value of the gravitational acceleration obtained by 
Bob is $g^+$ and, from Eq.~(\ref{eq.gratio}), we learn that Alice will
find that the gravitational acceleration in Bob's frame is $g^-$. It 
easy to conclude that every observer sitting at rest with the source of
gravity finds that the gravitational acceleration has the same value 
everywhere. The disagreement between observers sitting at a different
height can be interpreted as a consequence of local space-time dilation.

It worthwhile to note also that, moving along $x$ direction in an 
infinite uniform field, the local gravitational acceleration will
assume every possible value. This has an important consequence:
\begin{quote}
{\em All infinite extended uniform gravitational fields are equivalent.}
\end{quote}

This seems a strange conclusion, but remember that our starting point 
was purely ideal, and we did not analysed whether and how such a field
could be realized in the real world.
\section{Falling frame choice}
\label{space}
In this section I will discuss our FFRF choice, concluding that it is 
well motivated and indeed the only possible one.

\par After the identification of $\Sigma^-$ with the Rindler hyperbola 
$x^2-t^2=x_0^2$ (equation (\ref{eq.rin2}) in section \ref{acc}),
we could try to identify $\Sigma^+$ with a spatial translation of 
Eq.~(\ref{eq.rin2}):
\be 
\label{eq.trasl} (x- h)^{2}-t^{2}=x_0^2. 
\ee
\begin{figure}[hbp]
\begin{center}
\includegraphics[scale=0.15 ]{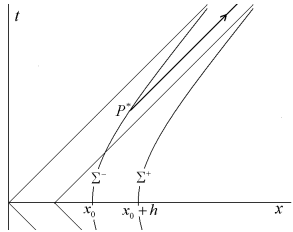}
\caption[short TOC caption]{\em Light ray emitted from $\Sigma^-$ in 
$P^*$ will never reach $\Sigma^+$} \label{fig.never}
\end{center}
\end{figure}
This choice would warrant the very same gravitational acceleration in 
$\Sigma^-$ and $\Sigma ^+$, but it gives rise to serious problems. The
most evident one is illustrated in Fig.~\ref {fig.never}, where a 
situation is exhibited in which $\Sigma^-$ cannot send a light signal that
reach $\Sigma ^+$. More generally it is possible to prove, also graphically, 
that the hyperbola $x^2-t^2=(x_0+h)^2$ (equation (\ref{eq.rioo})
of section~\ref{acc}) is the only one having constant radar distance from 
$\Sigma^-$.

\par Another important physical feature verified with our choice, 
is Lorentz invariance. For proving it, let us apply a boost $\Lambda$
of velocity $v= arcth \theta$ along $x$ direction to the hyperbola parametric equation
\cite{rindler,misner,mould} describing $\Sigma^-$
\be
\mathbf {\Lambda}=
\left(\begin{array}{ccc}
\cosh\theta  & -\sinh\theta  \\[5pt]
-\sinh\theta  & \cosh\theta
\end{array}\right)
\hspace{2cm}
\begin{cases}
&x=x_0\cosh g \tau\\
&t=x_0\sinh g \tau 
\end{cases}.
\ee
The trasformed Hyperbola
\be 
\label {eq.parametricboost}
\begin{cases}
&x'=x_0\cosh (g \tau-\theta)\\
&t'=x_0\sinh (g \tau-\theta)
 \end{cases},
\ee
shows that the only effect of a Lorentz boost is a shift in the proper
time origin $\Delta \tau=-\theta/g$.
%
%
\par The application of Lorentz boosts on hyperbolae relative to every possible 
height leaves their shape invariant and brings on the $x$ axis all events 
having the same velocity $v= arcth \theta$. This proves Lorentz 
invariance of our FFRF and gives also a further help in understanding 
simultaneity of events that are on the same straight line through the 
origin. On the other side, it is easy to prove that the family of hyperbolae 
obtained from Eq.~(\ref{eq.trasl}) is not Lorentz invariant.
\section{Metric Properties}
\label{metric}
In section \ref{measures} we calculated the time ratio between
two clocks at rest with the field
\be
\label{eq.timeratio}
\frac {\tau}{\tau_0}=1+gx,
\ee  
where now, in order to denote the height measured from a FFRF
(which starts the fall at $t=0$) we used $x$ instead of $h$.
Since the field we consider here is static, it is possible
to write the metric in diagonal form
\be
\label{eq.metric}
ds^2=A(x)dt^2-B(x)dx^2.
\ee
Writing the proper time elapsed for a clock at rest with the field
$d\tau=\sqrt{A(x)}dt$,
%
%
using Eq.~(\ref{eq.timeratio}) and comparing $d\tau$ with $d\tau_0$ 
is easy to find the explicit expression of the first metric term
\be
\label{adix}
A(x)=(1+gx)^2.
\ee
In order to find that of $B(x)$, it is possible to relate radar distances 
expressed using the metric coefficients with that we calculated 
in Eq.~(\ref{eq.radar-}). One has to solve the equation
\be
A(x_0)\int_{x_0}^x{\sqrt{\frac{B(x)}{A(x)}}}=\frac{1}{g}\log{(1+gx)}.
\ee
After setting $A(x)=1+gx$ and differentiating one finds:
\be
\frac{\sqrt{B(x)}}{1+gx}=\frac{1}{1+gx},
\ee
hence $B(x)=1$.
\par The above result can also be deduced using the general solutions of
the Einstein field equation for a flat space 
(imposing the condition $ R^{\mu}_{ \nu \sigma \rho}=0$).
Rohrlich was the first to show that in this case we have the following
relation between the metric coefficients \cite{ror}
\be
\label{eq.ror}
B(x)=\bigg ( \frac{1}{g}\frac{d}{dx}\sqrt{A(x)}\bigg )^2 
\ee
and, for $A(x)=(1+gx)^2$,  we find again $B(x)=1$. The resulting metric line element
\be
\label{eq.real}
ds^2=(1+gx)^2dt^2-dx^2
\ee
is deduced also in Mould's treatise \cite{mould}, in a formally different
way, but also based on the EP and on GR formalism (without involving
the field equation).
The Metric of Eq.~(\ref{eq.real}) is only one particular solution of the field 
equation, and it worths underlyng that without having discussed the direct EP
application it could not be possible to choose this solution between the infinity
that satisfy Eq.~(\ref{eq.ror}).
Rohrlich stressed that between these solutions three of them have a particular
physical interest \cite{ror}. 
The first one gives Eq.~(\ref{eq.real}), while the other two give the 
following metrics:
\be
\label{eq.metric3}
ds^2=(1+2gx')dt'^2-(1+2gx')^{-1}dx'^2
\ee
\be
\label{eq.metric2}
ds^2=e^{2gx''}(dt''^2-dx''^2),
\ee
which are easily obtained from Eq.~({\ref{eq.real}) using the
following coordinate transformations
\be
\label{eq.trasfgut}
1+gx=(1+2gx')^{1/2} \qquad t=t'
\ee
\be
\label{eq.trasfdes}
1+gx=e^{gx''} \qquad t=t''.
\ee
Analyzing the transformation of Eq.~(\ref{eq.trasfdes}) it is evident that, 
spatial coordinates used in the metric of 
Eq.~(\ref{eq.metric2}) are relative to an observer at rest with the field while,
as we showed in this section, the coordinate used in Eq.~(\ref{eq.real})  
describe measures performed from the FFRF.
\par Let us compare the above metrics with the Schwarzschild
line element (dropping the angular term)
\be
\label{eq.Schwarz1}
ds^2=\bigg (1-\frac{2GM}{r}\bigg )dt^2-\bigg (1-\frac{2GM}{r}\bigg )^{-1}dr^2.
\ee
\par This metric can be locally expressed with a first order approximation
\be
\label{eq.Schwarzapp}
A(x)=\bigg (1-\frac{2GM}{r}\bigg )\approx\bigg (1-\frac{2GM}{r_0}
+\frac{2GM}{r_0^2}(r-r_0)\bigg ).
\ee
With the substitution $x=(r-r_0)$, $g=2GM/r_0^2$ and translating the spatial origin
($gx'=gx-2GM/r_0$), Eq.~(\ref{eq.Schwarz1}) can be recast in the form 
\be
\label{eq.Schwarz2}
ds^2=(1+2gx')dt^2-(1+2gx')^{-1}dx'^2.
\ee
This local approximation must be used wiht care, having lost some of 
the most important properties of the Schwarzschild solution, the most
important of which is the Riemann curvature tensor $R^{\mu}_{\mu \nu \sigma \rho}$
that now has become zero everywhere, while in the Schwarzschild solution
we have only $R_{ \mu \nu}$=0.
Comparing with the homogeneous field metrics we find a full analogy with
Eq.~(\ref{eq.metric3}).
Analysing the transformation of Eq.~(\ref{eq.trasfgut}), we find the relevant thing
that the approximated Schwarzschild metric of Eq.~(\ref{eq.Schwarz2}) can be
obtained from the homogeneous field metric of Eq.~(\ref{eq.real}) with the only
request that the spatial factor became the Schwarzschild one ($B(x)=(1+gx')^{1/2}$).
From this fact one should argue that spatial lenght contraction is 
not due to the field strength only (or to its relative potential), 
but very probably to force variations present in a central gravitational field.
\section{Concluding remarks}
\label{conc}
As far I know, part of the analysis presented here is new, and hopefully 
could give some further physical insight on the problems at hand. 
My first intent in writing this paper was mainly pedagogical. 
For this reason I always tried to use simple mathematics, and sometimes
calculations are perhaps not performed in the shortest way.
\par We conclude the discussion with some general considerations. 
In the physical case of central Newtonian potentials there always
exists an asymptotic limit in which gravity vanishes, from where one
can imagine to let the fall of the FFRF start; unfortunately this cannot
be achieved in the case of uniform fields. 
Supposing the fall of an extended body to start from 
within a region in which the field does not vanish, and taking into 
account the fact that the light speed limit is valid for every signal, 
one find large differences depending on the way the body is released. 
For example, starting the body fall removing a support from its bottom, 
one has a `dilation' effect, because the top starts to fall later 
than the bottom, whereas releasing the hook from which it hangs, the 
falling body will experience a `contraction' during the fall. 
Our FFRF leads to a different rule: the fall starts 
simultaneously with respect to every local frame  and it is `programmed'
to have the same velocity along lines through the origin; observers 
at rest with the field will consider events on these lines simultaneous. 
Incidentally, with this choice, we have that the FFRF will measure 
gravity acceleration decreasing with height and this generate an 
asymptotic limit where gravity vanishes.
Unfortunately, this limit does not have a non-relativistic counterpart,
as it is in the case of Schwarzschild solution.
Some of these points are discussed in detail in Mould treatise \cite{mould}.
We finally remark that the approach used in this paper can also be usefull
when discussing the problem of a free falling charge in a gravitational 
background \cite{vallis}.  
\begin{acknowledgments}
I am very grateful to Silvio Bergia who suggested me to study clock's 
rates in a uniform gravitational field using only an accelerated free
falling frame. He gave me also moral support and useful suggestions 
during the preparation of this work. Stimulating discussions with Corrado
Appignani, Luca Fabbri and Fabio Toscano are also gratefully acknowledged.
\end{acknowledgments}
\end{document}